\documentclass[aps,pra,amsmath,amssymb,twocolumn,superscriptaddress,showpacs]{revtex4-1}
\usepackage{amsmath,amssymb}
\usepackage{graphicx}	
\usepackage{bm}

\def\vecb{\boldsymbol}
\def\be{\begin{eqnarray}}
\def\ee{\end{eqnarray}}

\begin{document}

\author{Shunsuke~A.~Sato}
\email{ssato@ccs.tsukuba.ac.jp}
\affiliation 
{Center for Computational Sciences, University of Tsukuba, Tsukuba 305-8577, Japan}
\affiliation 
{Max Planck Institute for the Structure and Dynamics of Matter, Luruper Chaussee 149, 22761 Hamburg, Germany}

\title{Two-step Brillouin zone sampling for efficient computation of electron dynamics in solids}

\begin{abstract}
We develop a numerical Brillouin-zone integration scheme for real-time propagation of electronic systems with time-dependent density functional theory. This scheme is based on the decomposition of a large simulation into a set of small independent simulations. The performance of the decomposition scheme is examined in both linear and nonlinear regimes by computing the linear optical properties of bulk silicon and high-order harmonic generation. The decomposition of a large simulation into a set of independent simulations can improve the efficiency of parallel computation by reducing communication and synchronization overhead and enhancing the portability of simulations across a relatively small cluster machine.
\end{abstract}

\maketitle

\section{Introduction \label{sec:intro}}

Recent development of laser technologies has enabled the study of light-induced electron dynamics in ultrafast and highly-nonlinear regimes \cite{RevModPhys.72.545,RevModPhys.81.163,MOUROU2012720}. In the ultrafast regime, attosecond electron dynamics in solids has become experimentally accessible, opening a novel avenue for the exploration of nonequilibrium electron dynamics induced by light \cite{doi:10.1126/science.1260311,doi:10.1126/science.aag1268,doi:10.1126/science.aan4737,Siegrist2019,doi:10.1063/5.0020649,Lucchini2021}. In the highly nonlinear regime, high-order harmonic generation induced by intense laser fields in solids has been intensively studied with the goal to develop a novel light-source and to establish a probe of material properties \cite{Ghimire2011,Schubert2014,Luu2015,Ghimire2019}. Laser processing is another important application of strong laser pulses and laser micromachining with femtosecond laser pulses via non-thermal processes have attracted much interest \cite{Gattass2008,Uteza2011,Balling_2013,Kanitz_2019,Salter2019}.

Despite the significance of these phenomena, understanding of their microscopic properties has still been under development because microscopic information of light-induced electron dynamics is difficult to experimentally access due to the complex nature of nonequilibrium and nonlinear electron dynamics in solids. An \textit{ab-initio} electron dynamics simulation based on time-dependent density functional theory (TDDFT) \cite{PhysRevLett.52.997} has been a powerful tool to investigate such highly-complex microscopic electron dynamics in solids \cite{SATO2021110274,doi:10.1063/1.4716192,PhysRevLett.118.087403,PhysRevB.92.205413}. However, the computational cost of such \textit{ab-initio} electron dynamics simulation tends to be large, and the simulation often requires a large supercomputer. Part of the origin of the large computational cost is the large $k$-point sampling requirement, which reflects the continuous nature of energy bands of solids.

In this work, we consider a two-step $k$-point sampling procedure to evaluate a large $k$-point sampling with a set of relatively small $k$-point samplings. This sampling procedure allows one to decompose a computationally heavy electron dynamics simulation into a set of relatively small electron dynamics simulations. As a result, each simulation can be executed with a smaller size of cluster machine, and the propagation of the whole electronic system can be evaluated by accumulating independent runs of a set of small simulations. Furthermore, the proposed two-step $k$-point sampling is almost perfectly parallelizable since each decomposed simulation can be executed independently. Therefore, the two-step procedure offers a novel degree of freedom for efficient computation of electron dynamics simulations in solids.

This paper is organized as follows. In Sec.~\ref{sec:method}, we first revisit electron dynamics simulation with TDDFT and introduce the two-step $k$-point sampling method. In Sec.~\ref{sec:result}, we examine the performance of the two-step sampling procedure in both linear and nonlinear regimes, using linear optical absorption and high-order harmonic generation in crystalline silicon. Finally, our finding is summarized in Sec.~\ref{sec:summary}. Hereafter, atomic units are used unless stated otherwise.

\section{Methods \label{sec:method}}

\subsection{Electron dynamics simulation \label{subsec:elec-dynamics}}

First, we briefly revisit an electron dynamics simulation based on TDDFT \cite{PhysRevLett.52.997,PhysRevB.62.7998}. The details of the method are described elsewhere \cite{SATO2021110274}. Light-induced electron dynamics in solids is described by the following time-dependent Kohn--Sham equation,
\be
i\frac{\partial}{\partial t}u_{b\vecb{k}}(\vecb r,t)=\hat h_{\vecb k}(t)u_{b\vecb{k}}(\vecb r,t),
\label{eq:tdks}
\ee
where $b$ is the band index, $\vecb k$ is the Bloch wavevector, and $u_{b\vecb k}(\vecb r,t)$ is the periodic part of the time-dependent Bloch orbital. The time-dependent Kohn--Sham Hamiltonian $\hat h_{\vecb k}(t)$ is given by
\be
\hat h_{\vecb k}(t) = \frac{\left [ \vecb p + \vecb k + \vecb A(t) \right ]^2}{2}
+ \hat v_{ion} + v_{Hxc}\left [ \rho(\vecb r,t) \right ],
\label{eq:ks-ham}
\ee
where $\vecb A(t)$ is a time-varying spatially uniform vector potential, which is related to an external laser electric field by $\vecb E(t)=-\dot{\vecb A}(t)$. The ionic potential is denoted as $\hat v_{ion}$ and may consist of non-local operators via the norm conserving pseudopotential approximation \cite{PhysRevLett.48.1425,PhysRevB.43.1993}. The Hartree-exchange-correlation potential is denoted as $v_{Hxc}\left [\rho(\vecb r,t) \right ]$ and is a functional of the electron density $\rho(\vecb r,t)$, defined as
\be
\rho(\vecb r,t) = \sum_b \frac{1}{\Omega_{BZ}}\int_{BZ} d\vecb k \left |u_{b,\vecb k}(\vecb r,t) \right|^2,
\label{eq:rho}
\ee
where $\Omega_{BZ}$ is the volume of the Brillouin zone. Note that all electronic orbitals $u_{b\vecb k}(\vecb r,t)$ are coupled with each other through the electron density $\rho(\vecb r,t)$ in the Hartree-exchange-correlation potential $v_{Hxc}\left [\rho(\vecb r,t) \right ]$.

\begin{widetext}
Once the time evolution of the Kohn--Sham orbitals are evaluated with Eq.~(\ref{eq:tdks}), one can compute physical quantities as a function of time with
\be
O(t) = \sum_b \frac{1}{\Omega_{BZ}} \int_{BZ} d\vecb k \int_{cell} d\vecb r u^*_{b,\vecb k}(\vecb r,t) \hat O 
u_{b,\vecb k}(\vecb r,t), 
\label{eq:observable}
\ee
where $O(t)$ is a physical quantity and $\hat O$ is the corresponding operator. For example, the current density as a function of time can be evaluated with
\be
\vecb J(t) = - \frac{1}{\Omega_{cell}}\sum_b \frac{1}{\Omega_{BZ}}\int_{BZ} d\vecb k \int_{cell} d\vecb r
u^*_{b,\vecb k}(\vecb r,t)  \vecb{\pi}_{\vecb{k}}(t) u_{b,\vecb k}(\vecb r,t),
\label{eq:current}
\ee
where $\Omega_{cell}$ is the volume of the cell and the kinetic momentum operator $\vecb{\pi}_{\vecb{k}}(t)$ is defined by
\be
\vecb{\pi}_{\vecb{k}}(t) = \frac{1}{i} \left [\vecb r, \hat h_{\vecb k}(t) \right ].
\ee

In practical calculations, the Brillouin zone integral in Eq.~(\ref{eq:rho}) and Eq.~(\ref{eq:observable}) is numerically evaluated as
\be
\frac{1}{\Omega_{BZ}} \int_{BZ} d\vecb k f(\vec k) \approx \frac{1}{N_1 N_2 N_3}\sum^{N_1}_{n_1=1}\sum^{N_2}_{n_2=1}\sum^{N_3}_{n_3=1} f(\vecb k_{n_1,n_2,n_3}),
\label{eq:bz-sampling}
\ee
where the integral is approximated by finite sampling of $k$-points. One of the most widely used approaches is Monkhorst--Pack sampling \cite{PhysRevB.13.5188}, which is given by
\be
\vecb k_{n_1,n_2,n_3} = \frac{2n_1 - N_1 -1}{2N_1}\vecb b_1 + \frac{2n_2 - N_2 -1}{2N_2}\vecb b_2 + \frac{2n_3 - N_3 -1}{2N_3}\vecb b_3,
\label{eq:mp-sampling}
\ee
where $\vecb b_j$ are the primitive reciprocal lattice vectors.
\end{widetext}

To obtain numerically converged results for the Brillouin zone integral Eq.~(\ref{eq:bz-sampling}), one needs to employ a sufficiently large number of sampling points. As will be demonstrated later, a huge number of $k$-points are required to obtain certain optical properties to reflect the continuous nature of the energy band of solids. Since all orbitals at different $k$-points are coupled with each other through the electron density $\rho(\vecb r,t)$, all of the states have to be propagated simultaneously. Hence, the computation of optical properties requires massive parallel computation for practical applications. However, such massive parallel computation is not usually efficient due to communication among different processes and their synchronization. Furthermore, larger supercomputers are often not easily accessible. Therefore, a decomposition of such a large simulation into a set of relatively small simulations is desirable from the viewpoints of computational efficiency and simulation portability.

\subsection{Two-step Brillouin zone sampling \label{subsec:two-step-bz}}

Here, we consider a method to decompose a large electron-dynamics simulation into a set of relatively small calculations. For this purpose, we evaluate the Brillouin zone integral of Eq.~(\ref{eq:bz-sampling}) with the following extended sampling:
\begin{widetext}
\be
\frac{1}{\Omega_{BZ}} \int_{BZ} d\vecb k f(\vec k) \approx \frac{1}{N_H}\sum^{N_H}_{m=1}\frac{1}{N_1 N_2 N_3}\sum^{N_1}_{n_1=1}\sum^{N_2}_{n_2=1}\sum^{N_3}_{n_3=1} f(\vecb k_{n_1,n_2,n_3,m}).
\label{eq:bz-two-step-sampling}
\ee
Here, the sampling $k$-points are defined as
\begin{equation}
  \vecb k_{n_1,n_2,n_3,m} = \frac{2n_1 - N_1 -1 +2q_m}{2N_1}\vecb b_1 + \frac{2n_2 - N_2 -1 +2p_m}{2N_2}\vecb b_2 + \frac{2n_3 - N_3 -1 +2r_m}{2N_3}\vecb b_3,
  \label{eq:shifted-k-points}
\end{equation}
where $(q_m, p_m, r_m)$ are uniform sample points in $(0,1)\times(0,1)\times(0,1)$ in $R^3$. To generate uniform sample points, one may employ a (quasi) random number generator. In this work, we employ the Halton sequence \cite{Halton1960} based on 2, 3, and 5 to generate uniform sample points, $(q_m, p_m, r_m)$.
\end{widetext}

Importantly, the integral of the left-hand-side of Eq.~(\ref{eq:bz-two-step-sampling}) can be exactly evaluated with the right-hand side in the large $N_H$ limit, even for finite numbers of $N_1$, $N_2$, and $N_3$, because the additional degree of freedom, $(q_m, p_m, r_m)$, densely samples the region discretized by the original Monkhorst--Pack sampling in Eq.~(\ref{eq:mp-sampling}). Therefore, one can evaluate the Brillouin zone integral by the two-step sampling by a coarse sampling first with $N_1 \times N_2 \times N_3$ $k$-points and a dense second sampling with (quasi) random numbers, $(q_m, p_m, r_m)$. 

In practical calculations, we employ the following two steps of the $k$-point sampling for two different purposes. The first sampling with $N_1\times N_2\times N_3$ $k$-points is employed to provide the converged time-dependent exchange-correlation potential $v_{Hxc}\left [ \rho (\vecb r,t) \right ]$ and the converged Kohn--Sham orbitals $u_{b\vecb k}(\vecb r,t)$. Once converged Kohn-Sham orbitals are obtained with the first sampling, one can accurately evaluate observables at each $k$-point as $f(\vecb k)=\sum_b \langle u_{b\vecb k}(t)|\hat O | u_{b\vecb k}(t)\rangle$. Then, the second sampling with $N_H$ points is employed to provide the converged observable with the Brillouin zone integral, Eq.~(\ref{eq:bz-two-step-sampling}). Hence, we propose the following two-step procedure for Brillouin zone sampling:
\begin{enumerate}
\item Provide $N_H$ uniform sample points $(q_m, p_m, r_m)$ in $(0,1)\times(0,1)\times(0,1)$.
\item For each sample point of $(q_m, p_m, r_m)$, further provide $N_1\times N_2\times N_3$ sample points with the shifted Monkhorst-Pack grid with Eq.~(\ref{eq:bz-two-step-sampling}).
\item For each sample point of $(q_m, p_m, r_m)$, independently perform TDDFT simulations with $N_1\times N_2\times N_3$ $k$-points and compute physical quantities $O_m(t)$.
\item Evaluate the average of all the TDDFT simulations as $O(t)=\sum^{N_H}_{m=1}O_m(t)/N_H$.
\end{enumerate}

The above procedure provides the converged observable $O(t)$ if $N_1$, $N_2$, and $N_3$ are large enough to provide converged Kohn--Sham orbitals and $N_c$ is large enough to provide the converged value of the Brillouin zone integral in Eq.~(\ref{eq:bz-two-step-sampling}). Within these conditions, one can decompose a large TDDFT simulation into relatively small simulations with $N_1\times N_2\times N_3$ $k$-points.

\section{Results and discussion \label{sec:result}}

In this section, we examine the performance of the two-step Brillouin zone sampling procedure introduced in Sec.~\ref{subsec:two-step-bz}. For this purpose, we consider two kinds of electron dynamics in solids. One is the linear response of solids to an impulsive distortion, while the other is the high-order harmonic generation under an intense light pulse. For both examples, we employ crystalline silicon. All calculations in this work have been performed with \textit{Octopus code}~\cite{doi:10.1063/1.5142502}. The primitive cell of crystalline silicon, which contains two silicon atoms, is employed. The real-space grid is discretized into $15^3$ grid points. For the description of electron-ion interaction, the Hartwigsen--Goedecker--Hutter pseudopotential is employed~\cite{PhysRevB.58.3641}. To describe electron--electron interaction, the adiabatic local density approximation is employed \cite{PhysRevB.45.13244}.

\subsection{Linear response calculation in time-domain \label{subsec:lin-res}}

First, we consider a linear response of solids to light. In practice, we compute the dielectric function of silicon with the real-time method \cite{PhysRevB.62.7998,SATO2021110274}. For this purpose, we employ the following impulsive distortion as a perturbation,
\be
\vecb A(t) = -k_0 \vecb e_{\beta} \delta(t),
\label{eq:impulse}
\ee
where $k_0$ is the strength of the impulsive field, and $\vecb e_{\beta}$ is a unit vector along the $\beta$-direction. Under this impulsive field, we compute the electron dynamics by solving the time-dependent Kohn--Sham equation, Eq.~(\ref{eq:tdks}), and evaluate the induced current $\delta \vecb J(t)$, defined as $\delta \vecb J(t)=\vecb J(t)-\vecb J(t<0)$ with Eq.~(\ref{eq:current}). The optical conductivity $\sigma_{\alpha \beta}(\omega)$ tensor is defined with the current and impulsive field as
\be
\tilde J_{\alpha}(\omega) = \sigma_{\alpha \beta}(\omega) k_0,
\ee
where $\tilde J_{\alpha}(\omega)$ is the $\alpha$-component of the Fourier transform of $\delta \vecb J(t)$,
\be
\tilde J_{\alpha}(\omega) = \int^{\infty}_0dt e^{i\omega t} W_L\left (\frac{t}{T_W} \right ) \vecb e_{\alpha}\cdot \delta \vecb J(t)
\label{eq:FT-current}
\ee
with the window function $W_L(x) = 1 - 3 x^2 + 2 x^3 $ in the domain $0<x<1$ and equal to zero outside. The duration of the window function, $T_W$, is set to $30$~fs. Furthermore, the dielectric function is given by
\be
\epsilon_{\alpha \beta}(\omega) = \delta_{\alpha \delta} + 4\pi i \frac{\sigma_{\alpha\beta}(\omega)}{\omega}.
\ee
Since silicon is an isotropic material, the dielectric function tensor is described by a single component as $\epsilon_{\alpha \beta}(\omega)=\epsilon(\omega)\delta_{\alpha \beta}$.

\begin{figure}[htbp]
  \centering
  \includegraphics[width=0.90\columnwidth]{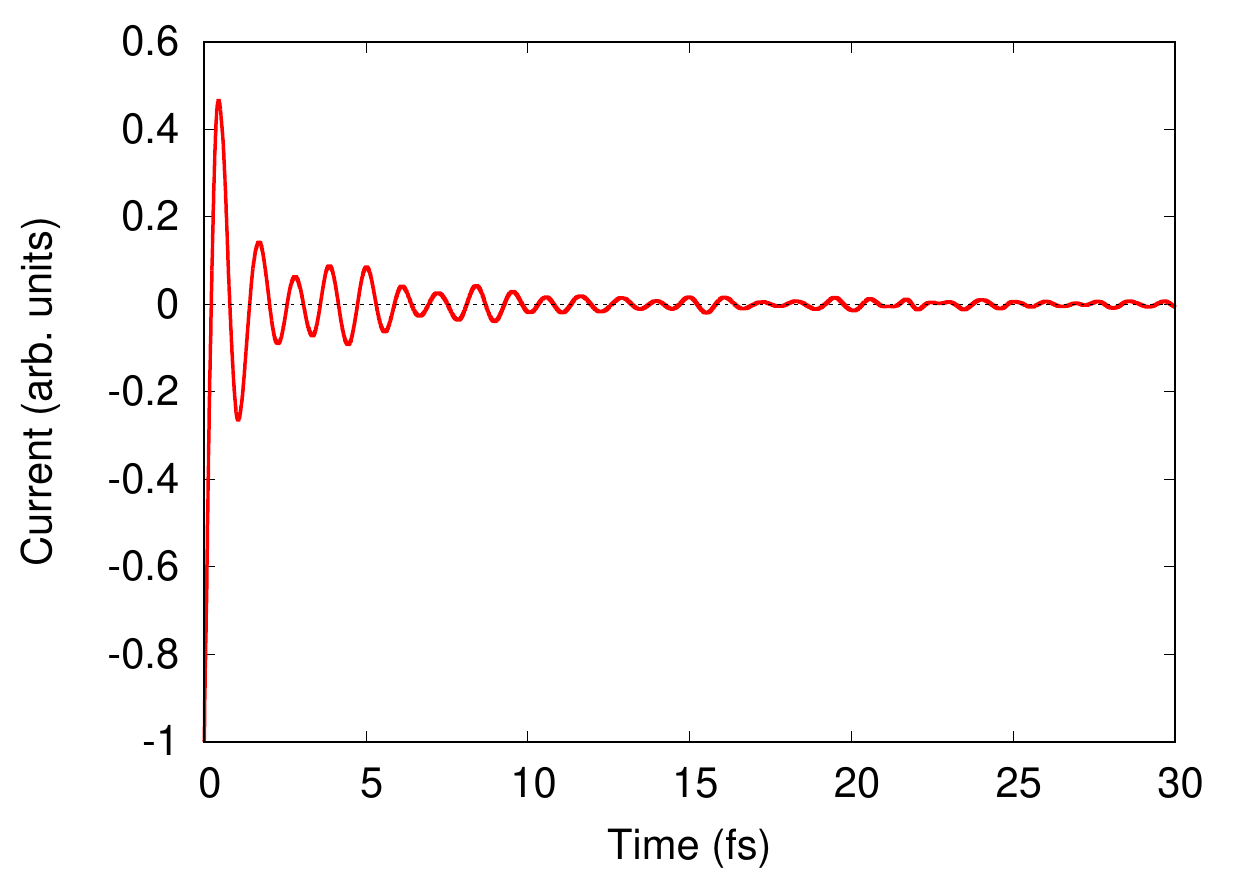}
\caption{\label{fig:current_lin}
Computed electric current in crystalline silicon induced by an impulsive distortion, Eq.~(\ref{eq:impulse}). The simulation is performed with Monkhorst--Pack sampling with $64^3$ $k$-points.
}
\end{figure}

As a reference to assess the two-step Brillouin zone sampling, we first evaluate the dielectric function of silicon with the original Monkhorst--Pack sampling, Eq.~(\ref{eq:mp-sampling}). Figure~\ref{fig:current_lin} shows the time-profile of the induced current, $\delta \vecb J(t)$, with the impulsive distortion, Eq.~(\ref{eq:impulse}). One sees that the current shows oscillatory behavior with damping. By applying the Fourier transformation, Eq.~(\ref{eq:FT-current}), to the current, we further evaluate the dielectric function.

\begin{figure}[htbp]
  \centering
  \includegraphics[width=0.90\columnwidth]{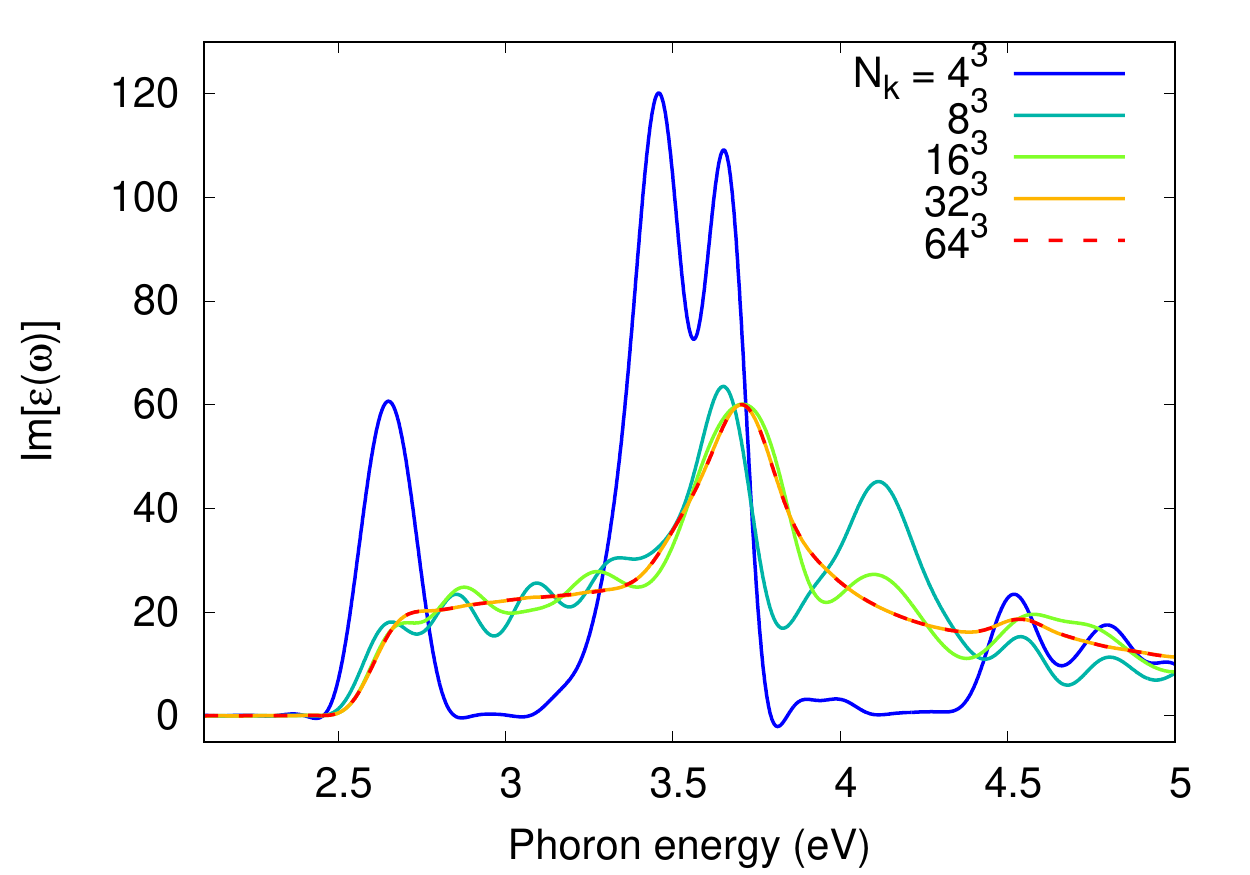}
\caption{\label{fig:eps_im_kconv}
Imaginary part of the dielectric function of silicon computed with the Monkhorst--Pack method. Results with different numbers of $k$-points, $N_k=N_1 \times N_2 \times N_3$, are shown.
}
\end{figure}

Figure~\ref{fig:eps_im_kconv} shows the imaginary part of the dielectric function of silicon computed with the Monkhorst--Pack method. The results with different numbers of $k$-points, $N_k=N_1\times N_2\times N_3$, are shown. One sees that the dielectric function becomes smoother for the larger number of $k$-points, and it is almost converged with $N_k=32^2$ $k$-points. On the other hand, the dielectric function becomes less smooth for a smaller number of $k$-points, reflecting the discretization of the continuous energy bands. Hence, a fine $k$-point sampling is indispensable to describe the continuity of the energy band, in addition to the convergence of each Kohn--Sham orbital. Assuming that the convergence of each Kohn--Sham orbital is realized with a relatively small number of $k$-points, we next apply the two-step Brillouin zone sampling to the linear response calculation.

To examine the two-step Brillouin zone sampling procedure, we fix the number of sampling points, $N_k=N_1\times N_2\times N_3$ in Eq.~(\ref{eq:mp-sampling}), to $8^3$ in this work unless stated otherwise. Note that $N_k$ should be large enough to provide well-converged Kohn--Sham orbitals. In contrast, $N_k$ does not need to be large enough to accurately evaluate the Brillouin zone integral since the fine $k$-point sampling in the Brillouin zone can be be addressed by the second sampling with the shift of $(q_m, p_m, r_m)$ in Eq.~(\ref{eq:bz-two-step-sampling}). In practical calculations, we first produce $N_H$ sets of (quasi) random numbers, $(q_m, p_m, r_m)$. We then perform the TDDFT simulation with $N_k$ $k$-points for each set of (quasi) random numbers and compute the induced current $\delta \vecb J_m(t)$ for each shift $(q_m, p_m, r_m)$ from $m=1$ to $N_H$. Finally, we average the current for different shifts as $\delta \vecb J(t)=\frac{1}{N_H} \sum^{N_H}_{m=1} \delta \vecb J_m(t)$ and evaluate the dielectric function with the above procedure.

\begin{figure}[htbp]
  \centering
  \includegraphics[width=0.90\columnwidth]{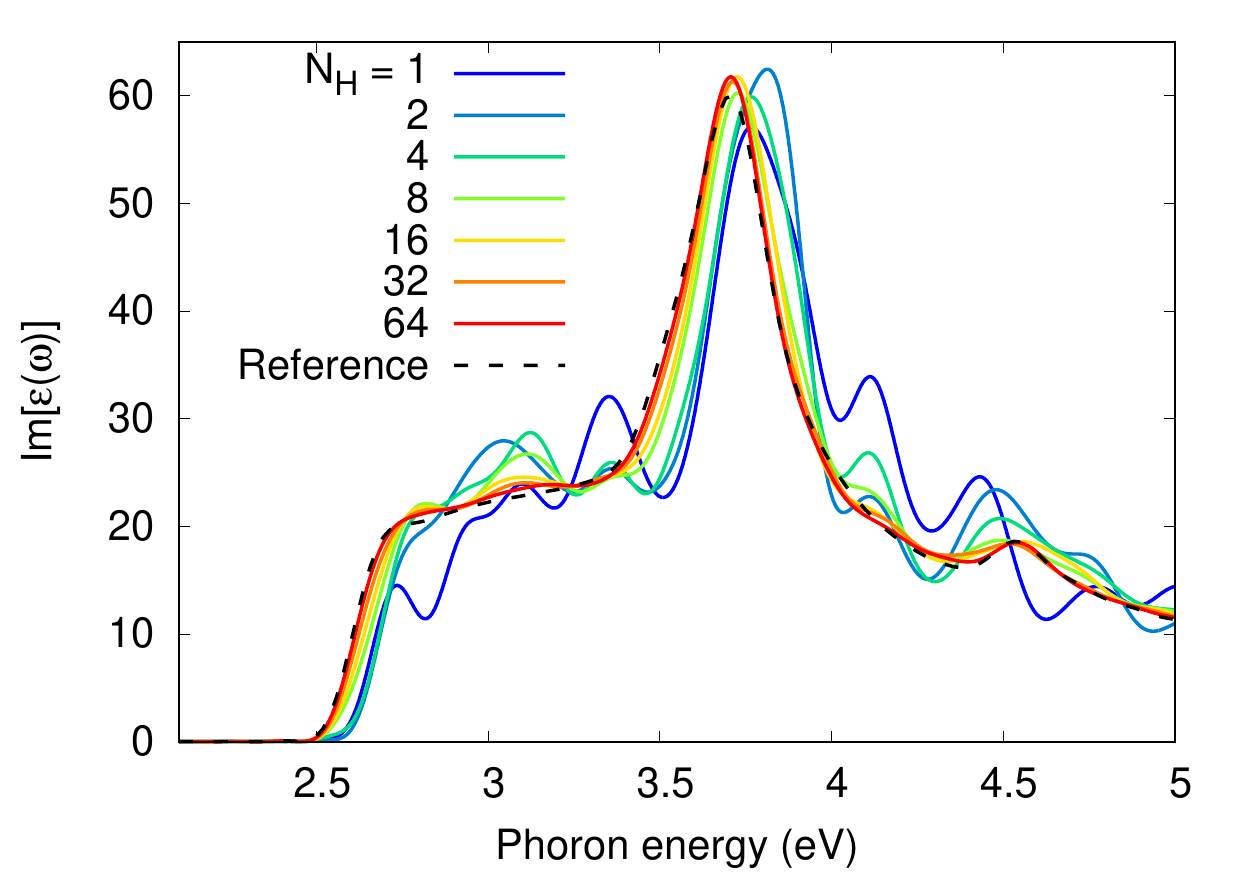}
\caption{\label{fig:eps_im_halton}
Imaginary part of the dielectric function of silicon computed with the two-step Brillouin zone sampling method. Results with different numbers for the second sampling, $N_H$, are shown. As a reference, the result of the Monkhorst--Pack method with $64^3$ $k$-points is also shown.
}
\end{figure}

Figure~\ref{fig:eps_im_halton} shows the imaginary part of the dielectric function computed with the two-step Brillouin zone sampling method. The results with different numbers of second-sampling points, $N_H$, are shown. Since the number of $k$-points of the shifted Monkhorst--Pack sampling $N_k$ is fixed to $8^3$, the total number of $k$-points in the two-step sampling method is given by $N_H\times N_k=8^3N_H$. As a reference, the result of the original Monkhorst--Pack method with $N_k=64^3$ $k$-points is shown as a black-dashed line in Fig.~\ref{fig:eps_im_halton}. As seen from Fig.~\ref{fig:eps_im_halton}, the results of the two-step sampling procedure are systematically improved by increasing $N_H$. In the present case, as seen from Fig.~\ref{fig:eps_im_halton}, $N_H=64$ is sufficient to provide a converged result identical to the original converged result in Fig.~\ref{fig:eps_im_kconv}. These results indicate that the present number of $k$-points for each TDDFT simulation, $N_k=8^3$, is sufficient to accurately describe the Hartree-exchange-correlation potential and time-dependent Kohn--Sham orbitals.

\begin{figure}[htbp]
  \centering
  \includegraphics[width=0.90\columnwidth]{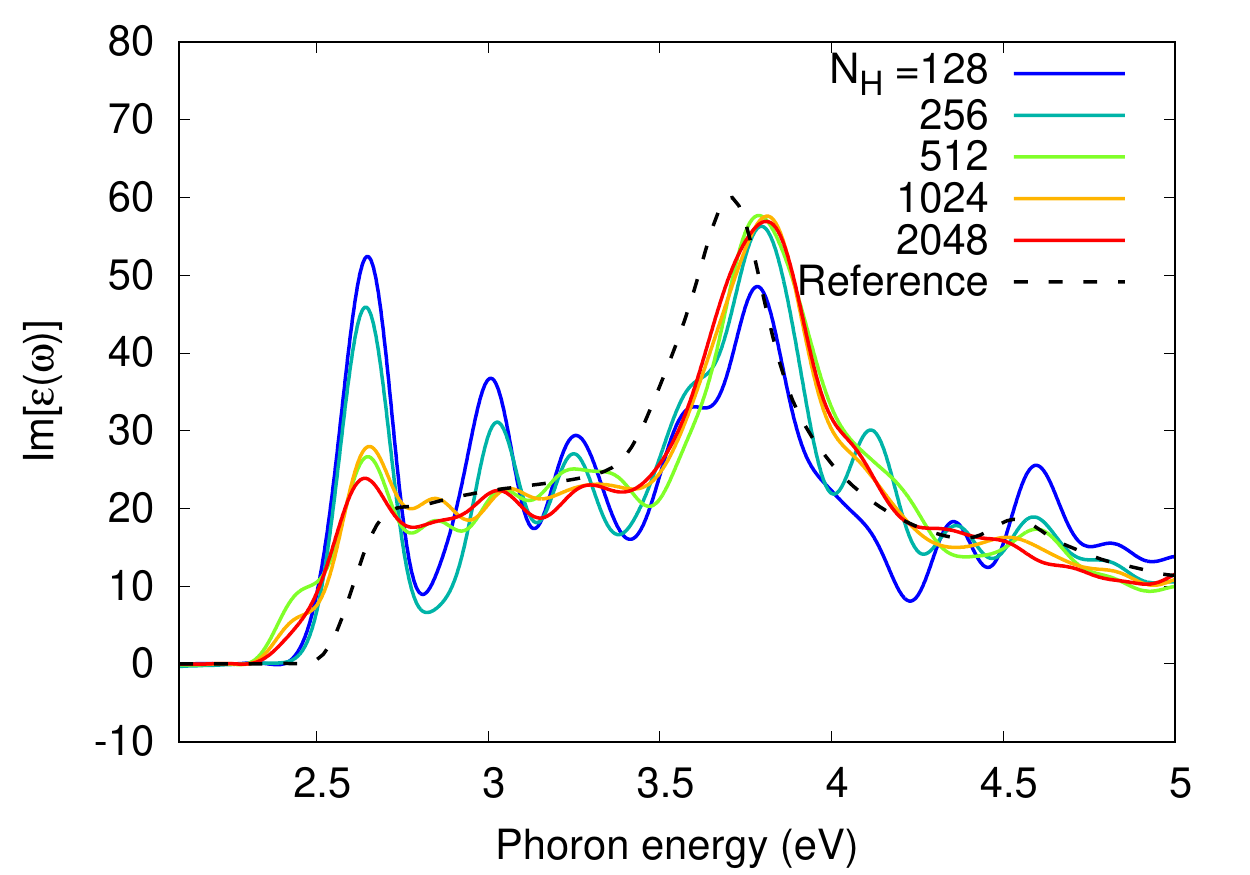}
  \caption{\label{fig:eps_im_halton_nk1}
Imaginary part of the dielectric function of silicon computed with the two-step Brillouin zone sampling method. Results with different numbers for the second sampling, $N_H$, are shown. The shaded area denotes the error bars that are evaluated as the standard error from $N_H$ data sets. As a reference, the result of the Monkhorst--Pack method with $64^3$ $k$-points is also shown.
}
\end{figure}

To demonstrate the importance of the number of sampling points, $N_k$, in Eq.~(\ref{eq:shifted-k-points}), we compute the dielectric function of silicon as performed in the above analysis but by fixing $N_k$ to $1$. Figure~\ref{fig:eps_im_halton_nk1} shows the computed results for different numbers of second-sampling points, $N_H$, but fixing $N_k$ to $1$. Here, the total number of $k$-points in the two-step sampling method is given by $N_H\times N_k=N_H$. As seen from Fig.~\ref{fig:eps_im_halton_nk1}, the results of the two-step sampling method do not converge to the converged result of the original method (black-dashed line) when $N_k$ is fixed to $1$. This indicates that the sufficiently large number of sampling points, $N_k$, has to be employed in the shifted Monkhorst--Pack sampling, Eq.~(\ref{eq:shifted-k-points}) to obtain correct results. If $N_k$ is not sufficiently large, the electron density and the Hartree-exchange-correlation potential are not well converged, and the resulting dynamics deviate from the correct dynamics. Therefore, one needs to check the convergence for $N_k$ in the two-step sampling method.

Note that the two-step Brillouin zone sampling method enables employment of a smaller number of $k$-points for each decomposed simulation and all simulations can be performed independently. In contrast, in the original method with the time-dependent Kohn--Sham equation, all $k$-points are coupled with each other, and all $k$-points have to be propagated simultaneously with synchronization. Therefore, the two-step Brillouin zone sampling method can be seen as the decomposition of a large TDDFT simulation into a set of relatively small simulations, offering merits from the viewpoint of computational efficiency. One of the most important merits is improvement of the parallel efficiency since decomposition into independent simulations reduces synchronization and communication among all $k$-points. Especially for a large simulation, synchronization and communication tend to occupy a large part of the total computational time. Hence, the two-step Brillouin zone sampling method can be used to reduce computational cost for large simulations. Another merit of the decomposition of a large simulation is the portability improvement of simulations across relatively small supercomputers and cluster machines. Thanks to the decomposition, relatively small simulations can be performed on a relatively small cluster machine. Furthermore, each simulation can be performed independently on each different machine. Hence, one can efficiently use various kinds of machines with the two-step sampling method and the portability of electron dynamics simulations can be enhanced.

\begin{figure}[htbp]
  \centering
  \includegraphics[width=0.90\columnwidth]{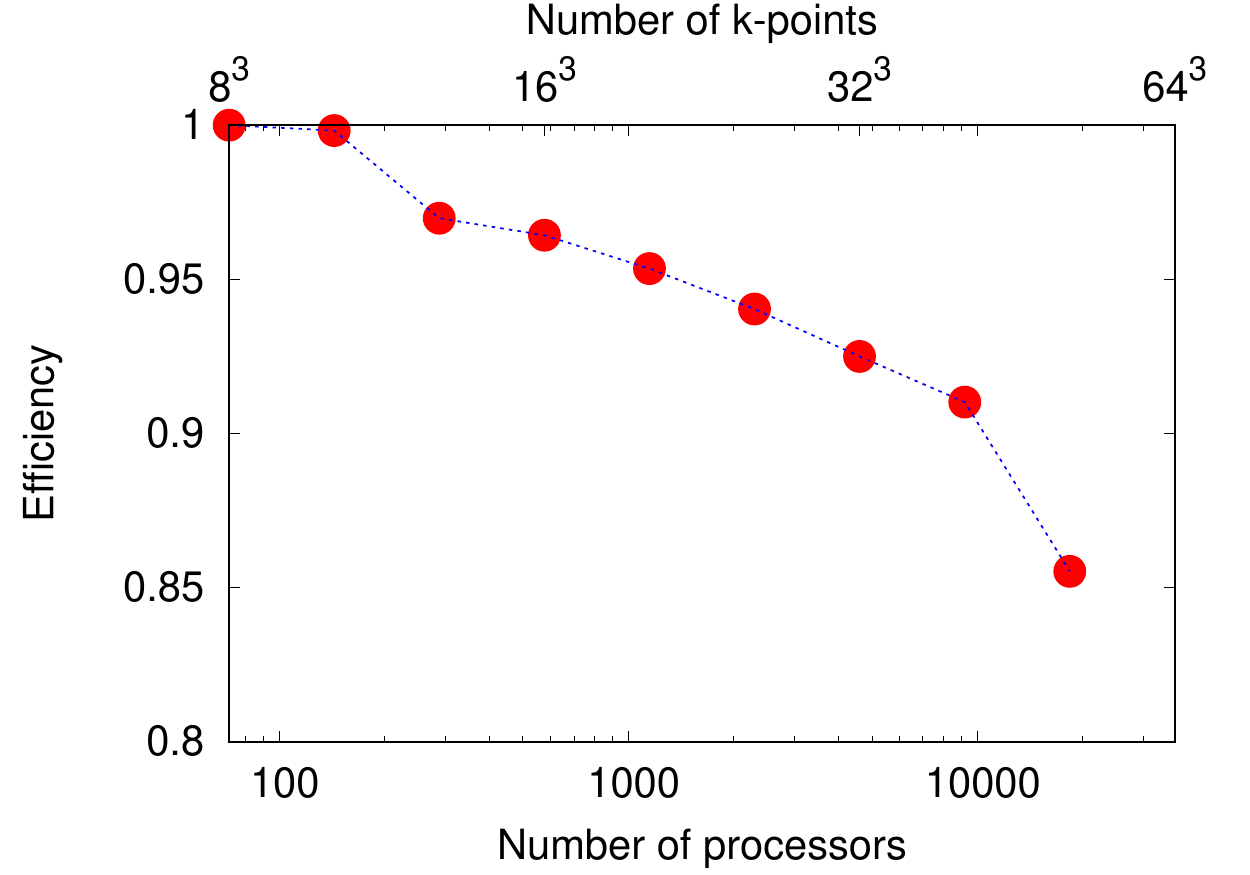}
\caption{\label{fig:performance}
Weak scaling performance of the electron dynamics simulation for crystalline silicon.
}
\end{figure}

To demonstrate the above merits of the two-step Brillouin zone sampling method, we evaluate the weak scaling efficiency for the parallel computation of the electron dynamics in solids. The weak scaling performance is analyzed by increasing both the number of processors and the problem size. By construction, the two-step Brillouin zone sampling method achieves the perfect scaling efficiency. Therefore, here we analyze the weak-scaling efficiency of the original $k$-point sampling method to demonstrate the improvement of the parallel efficiency by the two-step sampling method. The analysis was performed on the RAVEN HPC system of the Max-Planck Computing and Data Facility. For practical analysis, we evaluate the execution time of the electron dynamics simulation per time step by changing the number of processors in the parallel computation but fixing the ratio of the number of $k$-points and the number of processors. For the smallest simulation, we employ $8^3$ $k$-points in the electron dynamics simulation for crystalline silicon by using $72$ CPU cores. Figure~\ref{fig:performance} shows the evaluated weak scaling efficiency of the electron dynamics simulation with the original $k$-point sampling method as a function of the number of CPU cores (processors). Here, the efficiency is defined by the ratio of the execution time per time step for with $n_c$ CPU cores, $T_E(n_c)$, and that with $72$ CPU cores, $T_E(n_c=72)$: $\mathrm{Efficiency}=T_E(n_c=72)/T_E(n_c)$. As expected, the scaling efficiency decreases with the increase in the number of processors due to the communication and synchronization in the original method. By contrast, the two-step Brillouin zone sampling method offers the perfect weak scaling efficiency since a computationally large simulation can be decomposed into small independent simulations. Hence, the reduction of the efficiency of the original approach is equivalent to the improvement of the two-step sampling method. Furthermore, we note that the weak scaling efficiency cannot be evaluated for a larger problem size than $184320$ $k$-points in the present analysis since the number of required CPU cores exceeds the maximum number of available CPU cores for a general simulation job, which is $72~\mathrm{CPU~cores} \times 360~\mathrm{nodes}=25920$~CPU cores on the RAVEN HPC system. This indicates that if the maximum simulation run time, which is usually $24$~hours, is defined on a supercomputer, a large size problem may not be solved with the original $k$-point sampling method since all the $k$-points have to be propagated simultaneously. By contrast, even such a large simulation may be performed with the two-step Brillouin zone sampling method because the large simulation is decomposed into relatively small simulations, and the decomposed simulations can be executed independently as different simulation jobs. Therefore, the improvement of the portability of the two-step sampling method is also demonstrated here.

\begin{figure}[htbp]
  \centering
  \includegraphics[width=0.90\columnwidth]{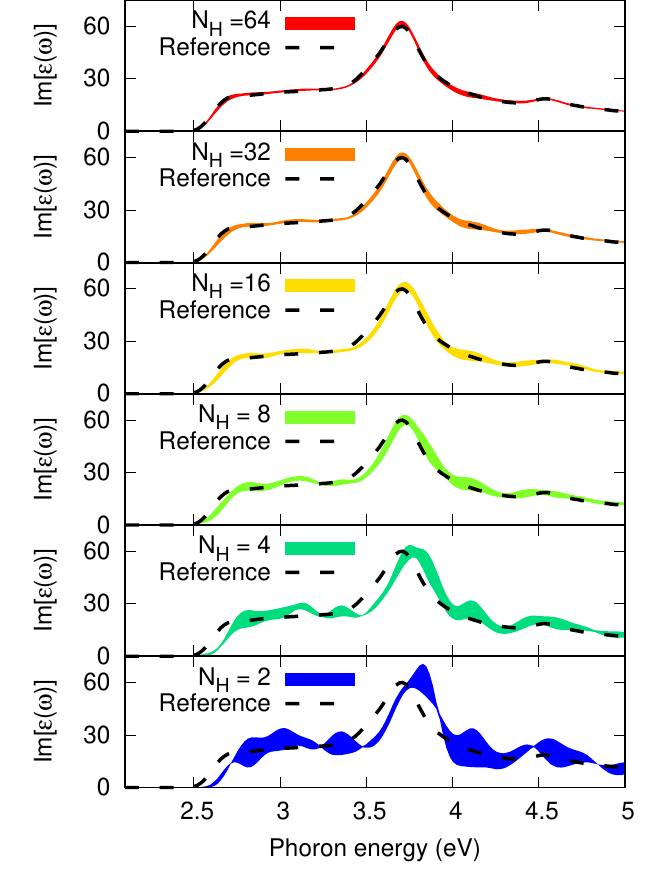}
  \caption{\label{fig:eps_im_halton_statistics}
Imaginary part of the dielectric function of silicon computed with the two-step Brillouin zone sampling method. Results with different numbers for the second sampling, $N_H$, are shown. The shaded area denotes the error bars that are evaluated as the standard error from $N_H$ data sets. As a reference, the result of the Monkhorst--Pack method with $64^3$ $k$-points is also shown.
}
\end{figure}

  In addition to the above merits from the viewpoint of computational resources, the two-step sampling method enables the efficient and straightforward convergence check for $k$-point sampling. For the case of the original method, as shown in Fig.~\ref{fig:current_lin}, the results of two different simulations must be compared. Furthermore, one of the simulations is always discarded, without being analyzed, since the other converged simulation is used for analysis. On the other hand, in the two-step sampling method, all simulation results have been integrated into the final results and no computational resources are wasted for the convergence check of $N_H$. Moreover, this feature of the two-step sampling method offers an additional degree of freedom to analyze the simulation results statistically. The results of the two-step sampling method such as Fig.~\ref{fig:eps_im_halton} can be seen as the average of $N_H$ sets of different simulation results. Hence, by analyzing $N_H$ data sets, the standard error of the mean, $\sigma_E$, of the results can be evaluated. Figure~\ref{fig:eps_im_halton_statistics} shows the imaginary part of the dielectric function of silicon computed with the two-step Brillouin zone sampling method for different numbers of second-sampling points, $N_H$, as shown in Fig.~\ref{fig:eps_im_halton}. While only the average values of the $N_H$ data sets are shown in Fig.~\ref{fig:eps_im_halton}, the shaded area in Fig.~\ref{fig:eps_im_halton_statistics} denotes the error bars for the mean. One can estimate the quality of the convergence of the calculation from the size of the error bars. In this way, the two-step sampling method offers a novel degree of freedom for better control of the convergence of the calculation. Note that, since the Halton sequence is employed in the present work, the above statistical analysis does not truly provide the standard error in the sense of Monte-Carlo simulation. However, one can simply employ random numbers instead of the Halton sequence and properly evaluate the standard error.

\subsection{High-order harmonic generation in solids \label{subsec:hhg}}

\begin{figure}[htbp]
  \centering
  \includegraphics[width=0.90\columnwidth]{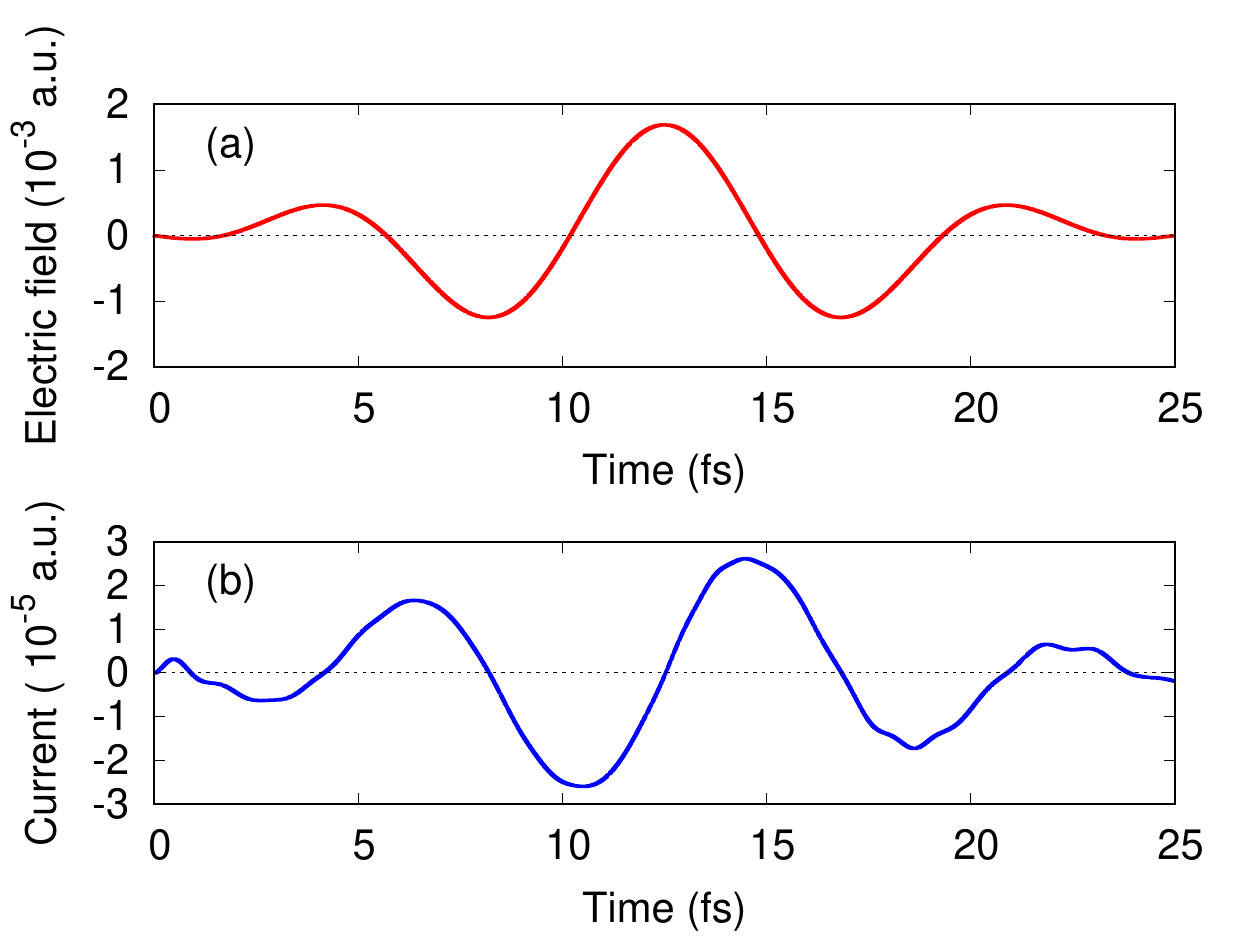}
\caption{\label{fig:current_hhg}
The computed current in silicon under the laser field given by Eq.~(\ref{eq:vec-pot-hhg}). The simulation is performed using the Monkhorst--Pack sampling with $64^3$ $k$-points.
}
\end{figure}

In the previous section, we examined the two-step Brillouin zone sampling method in the linear response regime. In this section, we extend analysis to a highly nonlinear regime. As an example of nonlinear phenomena, we consider high-order harmonic generation in silicon under an intense laser field. According to previous work \cite{PhysRevLett.118.087403}, we employ the following form of the vector potential as an external field
\be
\vecb A(t) = -\frac{E_0}{\omega_0}\vecb e_L \cos^4 \left (\pi \left (t-\frac{T_L}{2} \right ) \right ) \sin \left ( \omega_L \left (t-\frac{T_L}{2} \right ) \right ) \nonumber \\
\label{eq:vec-pot-hhg}
\ee
in the domain $0<t<T_L$ and equal to zero outside the domain. Here, $E_0$ is the peak field strength, $\omega_0$ is the mean frequency of the pulse, $\vecb e_L$ is the polarization direction, and $T_L$ is the full duration of the pulse. In this work, we set $E_0$ to $8.69$~MV/cm, $\omega_0$ to $413.3$~meV/$\hbar$, $\vecb e_L$ to the $\left (100 \right )$-direction, and $T_L$ to $25$~fs. Figure~\ref{fig:current_hhg}~(a) shows the time profile of the electric field $\vecb E(t)$ defined as $\vecb E(t)=-\frac{d}{dt} \vecb A(t)$.

We solve the time-dependent Kohn--Sham equation, Eq.~(\ref{eq:tdks}), with the vector potential given by Eq.~(\ref{eq:vec-pot-hhg}) and compute the electric current with Eq.~(\ref{eq:current}). Figure~\ref{fig:current_hhg}~(b) shows the $(100)$-component of the current induced by the field as a function of time, $\vecb J(t)$. We further evaluate the power spectrum of the emitted harmonics with the following Fourier analysis of the current
\be
I_{HHG}(\omega) = \omega^2 \left |
\int^{\infty}_{-\infty} dt e^{i\omega t} W(t)
\vecb e_L \cdot \vecb J(t)
 \right |^2,
\ee
where $W(t)$ is a window function given by
\be
W(t) = \cos^4 \left (\pi \left (t-\frac{T_L}{2} \right ) \right )
\ee
in the domain, $0<t<T_L$, and zero outside.

\begin{figure}[htbp]
  \centering
  \includegraphics[width=0.90\columnwidth]{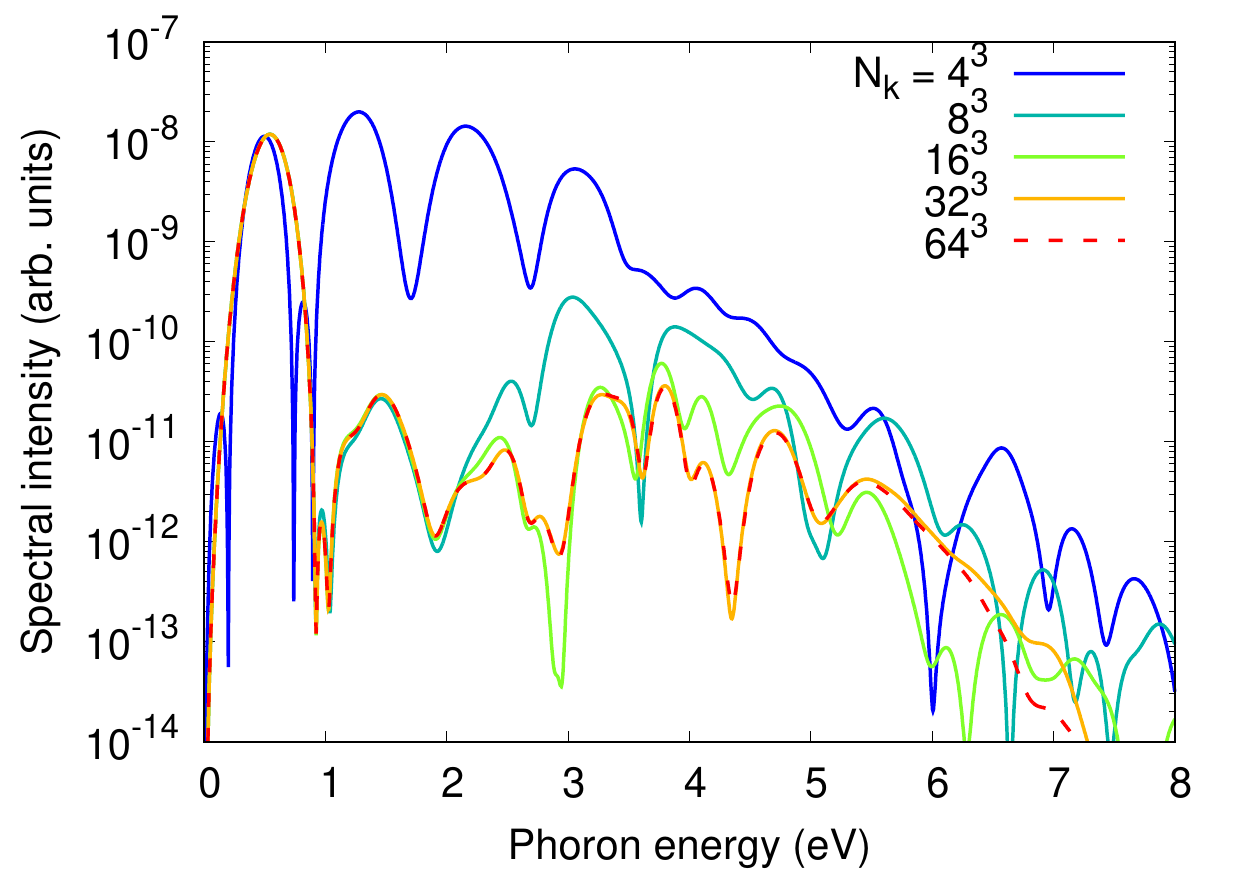}
\caption{\label{fig:hhg_kconv}
Computed spectrum of emitted harmonics with the Monkhorst--Pack method. The results for different numbers of $k$-points are shown.
}
\end{figure}

To examine the two-step Brillouin zone sampling method, we first study the convergence of the high-order harmonic generation with respect to the number of $k$-points for the original Monkhorst--Pack sampling, Eq.~(\ref{eq:mp-sampling}). Figure~\ref{fig:hhg_kconv} shows the computed spectra for different numbers of $k$-points. One sees that the intensity of the emitted harmonics is significantly overestimated for unconverged results with a smaller number $k$-points and the signal is reduced by a few orders of magnitude with the increase of the number of $k$-points. These results indicate that the emitted harmonics are significantly cancelled among various $k$-points and, hence, fine $k$-point sampling is indispensable for the evaluation of high-order harmonic generation. As seen from the figure, the well-converged result can be obtained with $32^3$ $k$-points with Monkhorst--Pack sampling.

\begin{figure}[htbp]
  \centering
  \includegraphics[width=0.90\columnwidth]{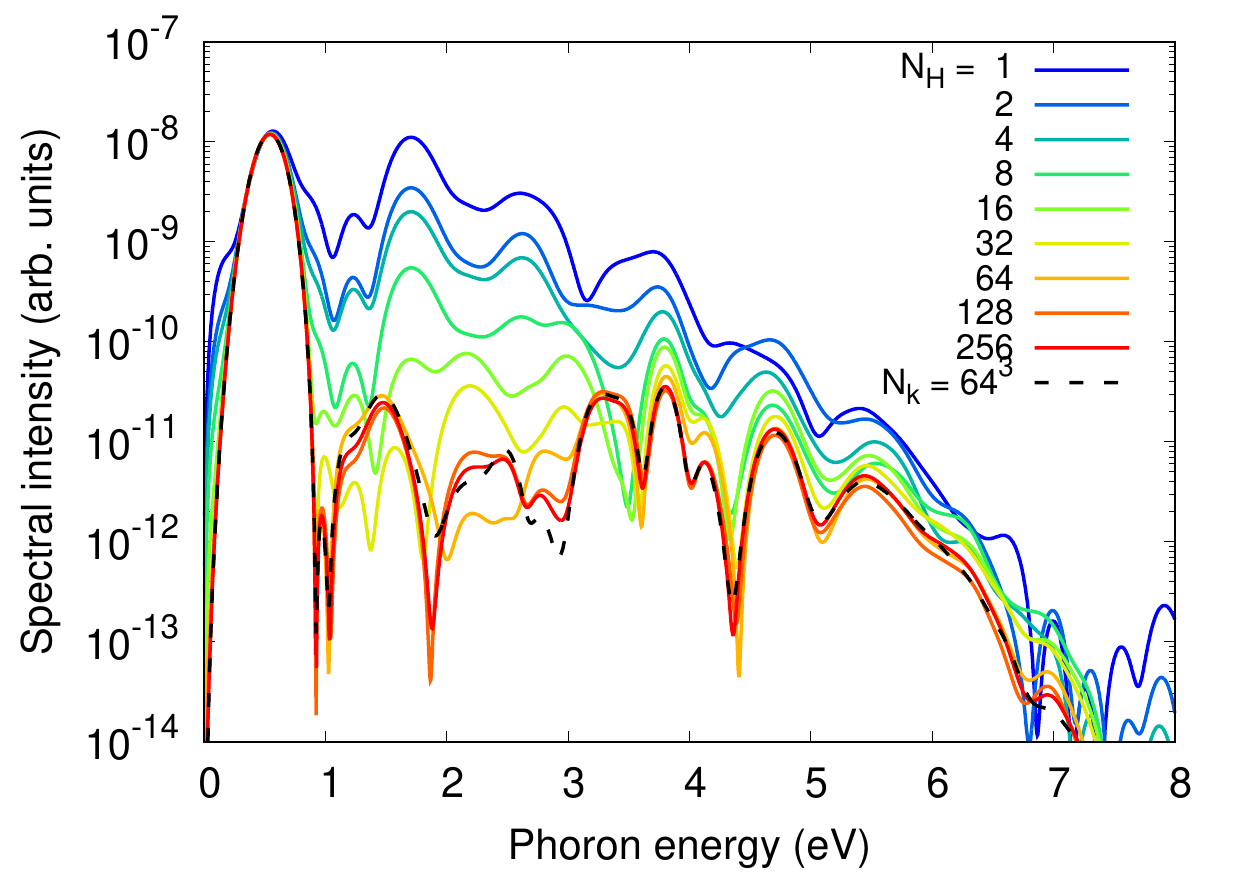}
\caption{\label{fig:hhg_halconv}
Computed spectrum of emitted harmonics with the two-step Brillouin zone sampling method. The results for different numbers of the second sampling, $N_H$, are shown. As a reference, the converged results of the Monkhorst--Pack method with $64^3$ $k$-points is also shown.
}
\end{figure}

We then study the convergence behavior of the two-step Brillouin zone sampling method. As described in the previous section, Sec.~\ref{subsec:lin-res}, we fix the number of first-sampling points to $N_k=N_1 \times N_2 \times N_3=8^3$ in Eq.~(\ref{eq:bz-two-step-sampling}). Then, we repeat the TDDFT simulation $N_H$ times with the shifted Monkhorst--Pack sampling, Eq.~(\ref{eq:shifted-k-points}). Finally, we evaluate the average of the induced current and compute the high-order harmonic generation spectrum from the averaged current. Figure~\ref{fig:hhg_halconv} shows the computed high-harmonics spectra with the two-step Brillouin zone sampling method. As a reference, the result of the original Monkhorst--Pack sampling with $64^3$ $k$-points is also shown as a black-dashed line. The results of the two-step Brillouin zone sampling method approach the converged result of the original Monkhorst--Pack sampling by increasing the number of secondary sampling, $N_H$. This result indicates that the significant cancellation of emitted harmonics among various $k$-points can be well described by the average of many independent runs of the TDDFT simulation with a relatively small number of $k$-points. Therefore, the validity of the two-step Brillouin zone sampling method has been demonstrated in the nonlinear regime.

\section{Summary \label{sec:summary}}

In conventional simulations for real-time electron dynamics with TDDFT, all Kohn--Sham orbitals are coupled with each other across all $k$-points via the Hartree-exchange-correlation potential, $v_{Hxc}[\rho(\vecb r,t)]$. Therefore, all orbitals must be propagated simultaneously. For optical responses of solids, one may need a very fine $k$-point sampling to accurately describe the continuity of the energy bands of solids and the computational cost of such simulations become large. For efficient computation of such simulations, we developed a Brillouin zone integration scheme for real-time propagation of electronic systems by decomposing a simulation with a large number of $k$-points into a set of simulations with a relatively small number of $k$-points, based on the two-step Brillouin zone sampling described in Eq.~(\ref{eq:bz-two-step-sampling}).

We examined the performance of the two-step Brillouin zone sampling method for both linear and nonlinear regimes, computing the linear optical property and high-order harmonic generation of silicon. In both regimes, the convergence of the two-step scheme to the converged results of the Monkhorst--Pack method has been demonstrated. These results indicate that the convergence of each Kohn--Sham orbital and Hartree-exchange-correlation potential can be achieved with a relatively small number of $k$-points while fine $k$-point sampling is required to compute the observables due to the continuity of energy bands. In the two-step sampling scheme, one can employ a relatively small number of $k$-point for the first sampling to obtain converged Kohn--Sham orbitals, while the fine $k$-point sampling can be performed by secondary sampling with (quasi) random numbers. Hence, the requirements of the convergence for both Kohn--Sham orbitals and continuity of energy bands can be efficiently realized.

The decomposition of a large simulation into a set of independent simulations can improve the efficiency of parallel computation since communication and synchronization overhead can be reduced. Furthermore, since small simulations may be executed on relatively small cluster machines, the simulation portability across various types of computers can be enhanced, enabling efficient use of computational resources with a novel degree of freedom.

\begin{acknowledgments}
This work was supported by JSPS KAKENHI Grant Numbers JP20K14382 and JP21H01842. We thank Enago for the English language review.
\end{acknowledgments}

\bibliography{ref}

\end{document}